\documentclass[aps,prl,twocolumn,groupedaddress]{revtex4}  
\usepackage{datetime}
\usepackage{amssymb}
\usepackage{graphicx}
\usepackage{bm}
\usepackage{amsmath}
\usepackage{epsfig}
\usepackage{float}
\usepackage{epstopdf}
\usepackage{fancybox}
\usepackage[section]{placeins}
\begin{document}
\title{Reply to Comment by Brown and Carrington on ``Phase-Space Approach to Solving the Time-Independent Schr\"{o}dinger Equation''}
 \author{Asaf Shimshovitz and David J. Tannor}
  \affiliation{Department of Chemical Physics, Weizmann Institute of Science, Rehovot, 76100 Israel}
\maketitle

The Comment of Brown and Carrington Jr. (BC) \cite{BrC15} has two main points: 1) that the contraction idea of Shimshovitz and Tannor (ST) can be used for any DVR basis, not necessarily periodic, and 2) that the biorthogonal basis introduced by ST is unnecessary.

On the first point we agree fully with BC.  Equation 4 of \cite{ShT12} reads:
\begin{equation}
\tilde{g}_n(x) = \sum_{m=1}^N \theta_m(x) g_n(x_m).
\end{equation}
In \cite{ ShT12}, the $\theta_m(x)$ were taken to be periodic sinc functions but this is a special case. As BC point out, the $\theta$ can be any DVR function, something that we discuss in \cite{TaT14} and applied in \cite{ShB14}.  We recognize that BC had this insight independently since \cite{TaT14} and \cite{ShB14} had not been published.

The realization that $\theta$ can be any DVR function seems at first glance to contradict the statement in \cite{ShT12} that ``By incorporating periodic boundary conditions into the vN lattice we solve a longstanding problem of convergence of the vN method."  To understand how these two seemingly contradictory statements can both be correct consider Fig. 1.  Note that the $\tilde{g}_n(x)$ defined using the periodic sinc $\theta$ functions are closely related to the original basis functions $g_n(x)$: they look similar, they are centered at the same lattice points in phase space as the $g_n(x)$, and all members of the basis are identical up to a shift in $x$ and $p$ (modulo the periodic boundary conditions).  Therefore, one is justified in saying that the use of the $\tilde{g}_n(x)$ converges the $g_n(x)$ basis.  In contrast, the use of a different set of $\theta$ functions, e.g. based on Legendre polynomials as used by BC or by us in
\cite{ ShB14}, will lead to a set of $\tilde g_n(x)$ that do not look particularly close to the original vN basis, will not be evenly spaced on a phase space lattice and will not be identical to each other.  One would be very hard pressed to call this ``convergence of the vN lattice" since it no longer bears any resemblance to the original vN lattice.  We note that there is an alternative method to converge the vN lattice that does not require periodicity \cite{PoS04}, but this is not what BC are discussing.

\begin{figure}
\begin{center}
\includegraphics [width=5.5cm]{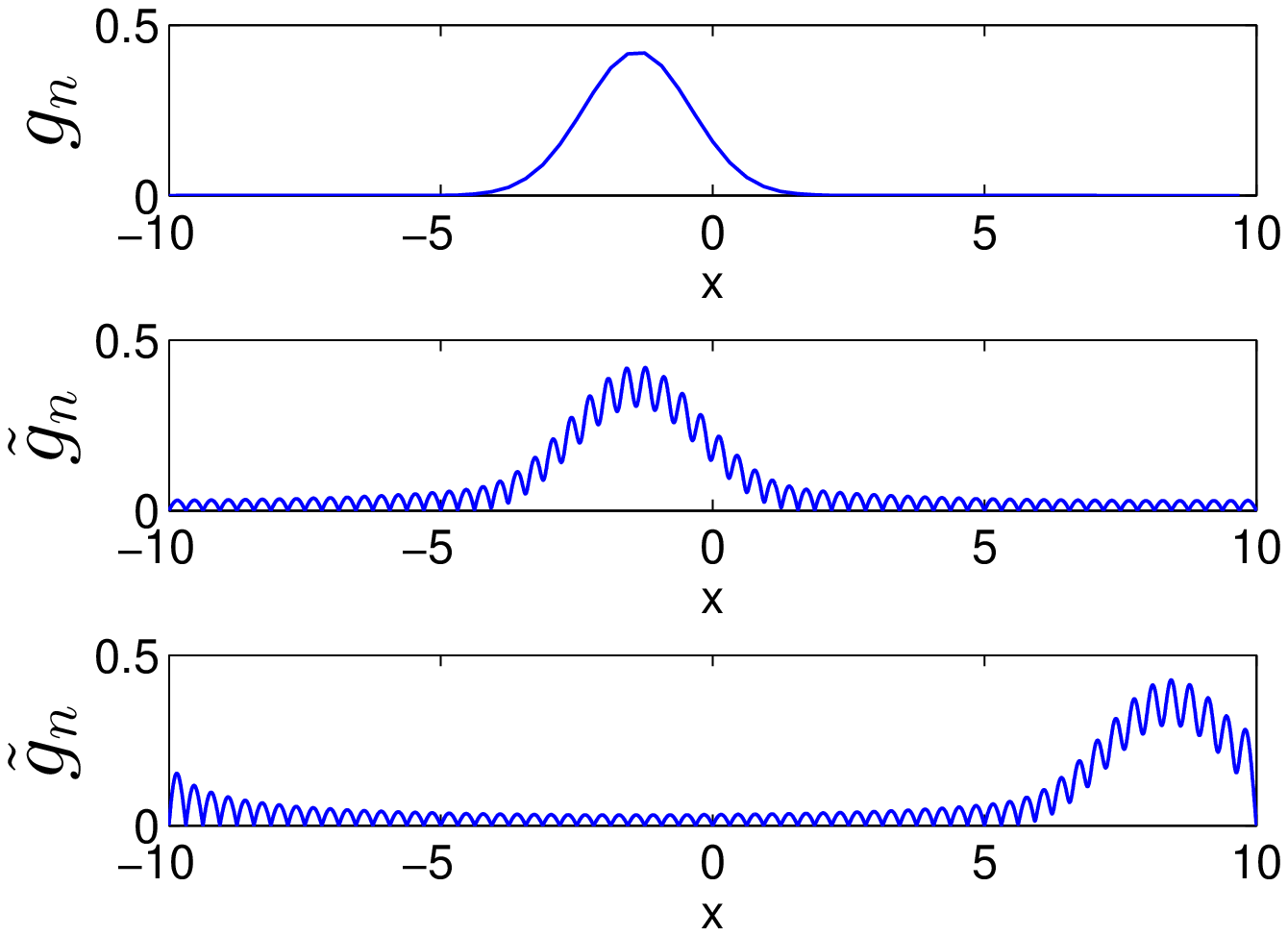}
\end{center}
\begin{center}
\vspace{-.6cm}
\caption{a) A typical member of the original vN basis, $g_n(x)$. b) A typical periodic vN basis function, $\tilde{g}_n(x)$; c) A periodic vN basis function near the boundary.}
\vspace{-1cm}
\label{g-tildeg}
\end{center}
\end{figure}
Turning to BCs claim that the biorthogonal basis is not required for pruning, we disagree. In BC's notation the biorthogonal basis is $\mathbf{S^{-1}G^\dag}$, which appears on both the left and right of $\mathbf{H}$ in their Eq. 5 where they are quoting ST. Although they claim there is no need for this basis, it appears on the left of $\mathbf{H}$ in their Eq. 6, so their real claim should be that the biorthogonal basis is needed on only one side of the representation of $\mathbf{H}$.

Finally, we disagree with BC's statement that: ``The simultaneous diagonalization (SD) basis of Refs. \cite{DaC05} was used in a similar fashion to contract a harmonic basis''. It is true that both the SD method and the pvb method create linear superpositions of 1-d functions, but there is no a priori reason to think that the SD functions are phase space localized.

Before closing, we point out some puzzling things in BC's Fig. 1b. 1) The pvb results without pruning should be identical to the results using a Finite Basis Representation with the same number of functions (compare the rightmost point in Fig. 2b in \cite{ShT12}). Thus, in Fig. 1b, one would expect that at $n_k=196$ the pink square should fall on the blue x and that the black dot should fall on the blue square. 2) The pink square at $n_k=196$ is actually orders of magnitude less accurate than its nearest neighbor, i.e. BC's Fig. 1b shows that pruning the pvb basis \emph{increases} the accuracy. 3) Even discounting these questions on Fig. 1b, note that it does not include a comparison with the pvb-FGH result from Fig. 1a. which is orders of magnitude more accurate.

This work was supported by the Israel Science Foundation and the Minerva Foundation.

\end{document}